\documentclass[twocolumn,showpacs,prl]{revtex4-1}
\usepackage{graphicx}
\usepackage{bm}
\begin{document}

\title{Restricted Thermalization for Two Interacting Atoms in a Multimode Harmonic Waveguide}
\author{V. A. Yurovsky}
\affiliation{School of Chemistry, Tel Aviv University, 69978 Tel Aviv, Israel}
\author{M. Olshanii}
\affiliation{Department of Physics, University of Massachusetts Boston, Boston MA 02125, USA}
\date{\today}

\begin{abstract}
In this article, we study the thermalizability of a system consisting of two atoms in a circular,
transversely harmonic waveguide in the multimode regime.
While showing some signatures of the quantum-chaotic behavior, the system fails to reach a thermal equilibrium 
in a relaxation from an initial state, even when the interaction between the atoms is infinitely strong.
We relate this phenomenon to the previously addressed unattainability of a complete quantum 
chaos in the \v{S}eba billiard [P.\ \v{S}eba, Phys. Rev. Lett., {\bf 64}, 1855 (1990)], 
and we conjecture the absence of a complete thermalization to be a generic property 
of integrable quantum systems perturbed by a non-integrable but well localized perturbation. 
\end{abstract}

% 05.70.Ln 	Nonequilibrium and irreversible thermodynamics 
% (see also 82.40.Bj Oscillations, chaos, and bifurcations in physical chemistry and chemical physics)
% 05.45.Mt 	Quantum chaos; semiclassical methods 
% 37.10.Gh 	Atom traps and guides
% 67.85.-d 	Ultracold gases, trapped gases (see also 03.75.-b Matter waves in quantum mechanics)
\maketitle

% =======================================================================

% Introduction

% =======================================================================

{\it Introduction}.--
The ultracold quantum gases have been long proven to be an ideal 
testbench for studying the fundamental properties of quantum systems. 
One of the most interesting and 
frequently addressed 
topics 
is the behavior of quantum systems in the vicinity of an 
integrable point.
Experimental results already include 
the suppression of relaxation of the momentum distribution 
\cite{kinoshita2006} and the modified decay
of coherence \cite{hofferberth2007,*hofferberth2008}.
Note that the fully developed quantum chaos with ultracold atoms 
has been  
investigated in experiments \cite{moore1995,*andersen2006}.

Traditionally, the underlying integrable system is represented 
by the Lieb-Liniger gas \cite{lieb1963} that is well suited to describe 
the dynamics of Van-der-Waals interacting bosons in a monomode
atomic guide (see \cite{yurovsky2008b} for a review). 
The nontrivial integrals of motion there are intimately 
related to both the one-dimensional character of the atomic motion
and to the effectively zero value of the two-body interaction range.
The causes for lifting integrability include the virtual 
excitation of the transverse modes during the collision
\cite{mora2004,*yurovsky2006,*mazets2008} and the coupling between 
two parallel weakly interacting Lieb-Liniger gases \cite{hofferberth2007,*hofferberth2008}.

One would expect that in the {\it multimode regime}, when the motion becomes 
substantially multidimensional, no remnants of the former integrals 
of motion will survive. However, in this article we show that for the 
case of only two interacting atoms a strong separation between the 
transverse and longitudinal degrees of freedom remains, even in the case 
of the infinitely strong interaction between the atoms. 
We attribute this effect to
the short-range nature of the interaction.

For our system we study both the degree of the eigenstate thermalization 
and the actual thermalization in an expansion from a class of realistic initial 
states. The eigenstate thermalization \cite{shnirelman1974,*feingold1986,
*flambaum1997,*deutsch1991,srednicki1994,rigol2008}
-- the suppression of the 
eigenstate-by-eigenstate variance of quantum expectation values 
of simple observables -- provides an ultimate upper bound for 
a possible deviation of the relaxed value of an observable 
from its thermodynamical expectation, for any initial state in principle. 
However, recently a new direction of research has emerged:
quantum quench 
in many-body interacting systems \cite{kinoshita2006,hofferberth2007,*hofferberth2008,rigol2008,berman2004,*calabrese2007,*flambaum2001,*kollath2007,
*manmana2007,*sengupta2004,*polkovnikov2008}.
In this class of problems, the initial state is inevitably decomposed into a large superposition
of the eigenstates of the Hamiltonian governing the dynamics, and the discrepancy between 
the result of the relaxation and thermal values is expected to be diminished.
In our article we show that in the case of two short-range-interacting atoms 
in a harmonic waveguide, a complete thermalization can never be reached under either scenario. 

We relate the absence of thermalization in our system
to the previously addressed unattainability of a complete quantum chaos
in the \v{S}eba billiard -- flat two-dimensional rectangular 
billiard with a zero-range scatterer in the middle \cite{seba1990}.
Similarly to our system, the \v{Seba} billiard does show some signatures of the
quantum-chaotic behavior, although 
both the level statistics \cite{seba1991,albeverio1991,bogomolny2001}
and the momentum distributions in individual eigenstates \cite{berkolaiko2003} 
show substantial deviations from the quantum chaos predictions.
Both the \v{S}eba billiard (first suggested and solved in Ref.\ \cite{seba1990}) 
and our system allow for an exact analytic solution,
in spite of the absence of 
a complete set of
integrals of motion.

% =======================================================================

% The system

% =======================================================================

{\it The system}.--
Consider two short-range-interacting atoms in a circular, transversally harmonic waveguide. 
In this case the center-of-mass and the relative motion 
can be separated. 
Note that the harmonic approximation for the transverse confinement has been proven 
to model real life atom waveguides with high accuracy. Predictions of the analogous 
model involving an infinite waveguide \cite{olshanii1998,bergeman2003} has been confirmed 
by a direct experiment-theory comparison of the energies of the quasi-one-dimensional dimers
\cite{moritz2005} and numerical calculations involving anharmonic potentials 
(see review \cite{yurovsky2008b} for references).

The unperturbed Hamiltonian of the relative motion is given by the sum of the longitudinal and  transverse
kinetic energies and the  transverse trapping energy $\hat{U} = \mu\omega_{\perp}^2\rho^2/2$,
\begin{eqnarray}
\hat{H}_0 = -{\hbar^2 \over 2\mu}{\partial^2\over\partial z^2}
-{\hbar^2\over 2\mu} \Delta_{\rho}+ \hat{U} - \hbar\omega_{\perp}
\quad,
\label{H_2b}
\end{eqnarray}
where 
$\omega_{\perp}$ is the transverse frequency, 
$\Delta_{\rho}$ is the transverse two-dimensional Laplacian,
and $\mu$ is the reduced mass.
In our model, the 
transverse and longitudinal degrees of freedom are coupled
by a potential of a Fermi-Huang type \cite{yurovsky2008b}:
\begin{equation}
\hat{V} = {2\pi\hbar^2 a_{s}\over \mu}\delta_{3}({\bm r}) (\partial/\partial r)(r\,\cdot ),
\label{Fermi-Huang}
\end{equation}
were, $a_{s}$ is the three-dimensional $s$-wave scattering length.
This approximation is valid at low collision energies (e.g., below 300 mK for Li, see a recent review \cite{chin2009},
while the ultracold atom energies lie well below 1 $\mu$K). 
The model (\ref{Fermi-Huang}) stands 
in a excellent agreement with the experimentally measured equation of state of the waveguide-trapped Bose gas
\cite{kinoshita2004} and with numerical calculations with finite-range potentials \cite{bergeman2003,naidon2007}.

The ring geometry of the waveguide imposes period-$L$ boundary conditions along $z$-direction. 
In what follows, we will restrict the Hilbert space to 
the states that have zero $z$-component of the angular momentum and that are even under 
the $z \leftrightarrow -z$ reflection;
the interaction has no effect on the rest of the Hilbert space. 
The unperturbed spectrum is given by (see \cite{supplement})
%%
%\begin{equation}
%%
$
E_{nl} = 2 \hbar \omega_{\perp} n + \hbar^2 (2\pi l/L)^2/(2\mu)
$,
%\label{E0_2b}
%%
%\end{equation}
%%
where $n \geq 0$ and $l\geq 0$ are the transverse and longitudial quantum numbers, respectively.

The interacting eigenfunctions with the eigenenergy $E_{\alpha}$ can be expressed as
(see \cite{supplement})
\begin{eqnarray}
\langle \rho ,z | \alpha\rangle =
C_{\alpha} \sum\limits^{\infty
 }_{n=0}{\cos\left( 2\sqrt{\epsilon _{\alpha}-\lambda n}\zeta \right)
 \over \sqrt{\epsilon _{\alpha}-\lambda n}\sin\sqrt{\epsilon _{\alpha}-\lambda
 n}} e^{-\frac{\xi}{2}} L^{(0)}_{n}\!\left(\xi\right) 
,
\label{psieps}
\end{eqnarray}
where the rescaled energy $\epsilon_{\alpha}$ is given by $\epsilon_{\alpha}\equiv\lambda E_{\alpha}/\left(
 2\hbar\omega_{\perp }\right)$, $\lambda \equiv\left( L
/a_{\perp }\right) ^{2}$  is the aspect ratio,
$\zeta \equiv  z/L-1/2$, \,$\xi \equiv \left(\rho/a_{\perp }\right)^{2}$, \,   
$L^{(0)}_{n}(\xi)$ are the Legendre polynomials, 
$a_{\perp }=\left(\hbar/\mu\omega _{\perp}\right)^{1/2}$ is the size of the 
transverse ground state, and 
$C_{\alpha} $ is the normalization constant.
The eigenenergies are solutions of the
following transcendental equation (see \cite{supplement}):
\begin{equation}
\sqrt{\lambda}\sum\limits^{\infty }_{n=0}
{\cot\sqrt{\epsilon_{\alpha} -\lambda n}+i\over \sqrt{\epsilon_\alpha
 -\lambda n}}-\zeta \left( {1\over 2},-{\epsilon_{\alpha} \over \lambda
 }\right) ={a_{\perp }\over a_{s}} 
\quad, 
\label{Eqeps}
\end{equation}
where $\zeta \left( \nu ,x\right) $ is the Hurwitz $\zeta $-function
(see \cite{yurovsky2008b}).
Similar solutions were obtained for
two atoms with a zero-range interaction in a cylindrically-symmetric
harmonic potential \cite{Idziaszek2005} (that system was analysed numerically
in Ref.\ \cite{bolda2003}).
Higher partial wave scatterers were analyzed in the Ref.\ \cite{Kanjilal2007}.

At rational values of $\lambda/\pi^2$, 
the unperturbed energy spectrum shows degeneracies that
are not fully lifted in the full [deduced from (\ref{Eqeps})] spectrum.
To minimize the effect of the degeneracies, we,
following Ref.\ \cite{seba1990},
fix the length of the guide to $(L/a_{\perp})^2 \equiv \lambda = 2 \phi_{gr} \pi^7 \approx 9774$,
where $\phi_{gr} = (1+\sqrt{5})/2$ is the golden ratio. 

% =======================================================================

% Tests

% =======================================================================

{\it Standard quantum-chaotic tests}.--
The results of 
the study of the level spacing distribution 
in our system 
are fully consistent with the analogous results for the  
\v{S}eba billiard \cite{seba1991,albeverio1991}.
For the energy range $E \gtrsim 100 \hbar \omega_{\perp}$
and the aspect ratio $\left(L/a_{\perp }\right)^{2} = 2 \phi_{gr} \pi^7$,
the distribution quickly converges to 
the \v{S}eba distribution \cite{seba1991} at $a_{s} \gtrsim 10^{-1}a_{\perp }$. 
This distribution does show a gap at small level spacings but
fails to reproduce the Gaussian tail predicted by the
Gaussian Orthogonal Ensemble.
At small $a_{s}$ the level spacing distribution tends to
the Poisson one. 
We have also verified that in the unitary regime, $a_{s} \gg a_{\perp }$, the statistics of the point-by-point variations 
of the eigenstate wavefunction in the spatial representation is close to the Gaussian one,
according to the general prediction \cite{Berry1977} and a particular observation 
in the case of the \v{S}eba billiard \cite{seba1990}.

% =======================================================================

% ETH

% =======================================================================

\begin{figure}
\includegraphics*[width=3.5in]{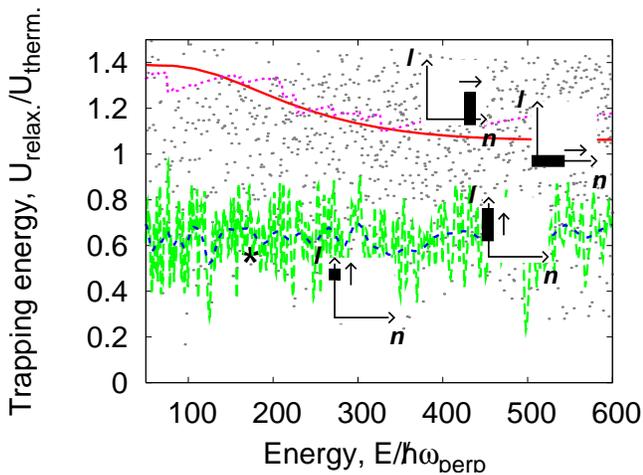}
\caption
{
The time average of the transverse trapping energy
$U_{\mbox{\scriptsize relax.}} \equiv \lim_{t_{max}\to\infty} (1/t_{max}) 
\int_{0}^{t_{max}}\! \langle \psi(t) | \hat{U} | \psi(t) \rangle$
after relaxation from an initial state as a function
of the energy of the initial state $E$ [see Hamiltonian (\ref{H_2b})]. Four families 
of the initial states are considered. Long-dashed (green) line:
state (\protect\ref{pdn}) with $n_{0}=0$, $\delta=0.99$,
and the energy being controlled via scanning $l_{0}$;
short-dashed (blue) line: the same as the previous one, except 
for $\delta=.1$; dotted (purple) line: the state (\protect\ref{pdn})
with $\delta=0.1$ and $l_{0}=0$ and the energy control via $n_{0}$; 
solid (red) line: state (\protect\ref{pdkk}) with
$\kappa_{2}=2\kappa _{1}$, $\delta=0.99$, $l_{0}=0$ and 
the energy control via $\kappa_1$. Legends schematically show 
the distribution of the initial states over the quantum numbers;
arrows indicate the direction in which the parameters controlling 
the energy are scanned. 
The asterisk (*) corresponds to the 
set of parameters used to produce the Fig.\ \ref{Fig_pdnm}.
Grey dots show the quantum expectation value of
the trapping energy $U$ for every hundreds eigenstate of the
full Hamiltonian (\ref{H_2b},\ref{Fermi-Huang}).
All the values of the trapping energy are shown with respect to its 
microcanonical expectation value $U_{therm.} \approx E/3$. 
For all the points, $\left(L/a_{\perp }\right)^{2} = 2 \phi_{gr} \pi^7 \approx 9774$ 
and $a_{s} = 10^{6} a_{\perp}$.
}
\label{Fig_csie}
\end{figure}

{\it Eigenstate thermalization}.--
According to the eigenstate thermalization hypothesis \cite{shnirelman1974,*feingold1986,
*flambaum1997,*deutsch1991,srednicki1994,rigol2008},
the ability of an isolated quantum system (with few or many degrees of freedom regardless) to thermalize 
follows from the suppression of the 
eigenstate-by-eigenstate fluctuations of the quantum expectation values of relevant observables.
Scattered dots at the Fig.\ \ref{Fig_csie} show the 
quantum expectation values of the transverse trapping energy, $\langle \alpha | \hat{U} | \alpha \rangle$
as a function of the eigenstate energy.
The variation of $\langle \alpha | \hat{U} | \alpha \rangle$ remains 
(in contrast to the quantum-chaotic billiards \cite{barnett2006})
of the order of the mean, even for the energies much larger than any conceivable energy scale of the system. 
Here and below, we work in the regime of the infinitely strong interactions, $a_{s} = 10^{6} a_{\perp}$. 

\begin{figure}
\includegraphics*[width=3.5in]{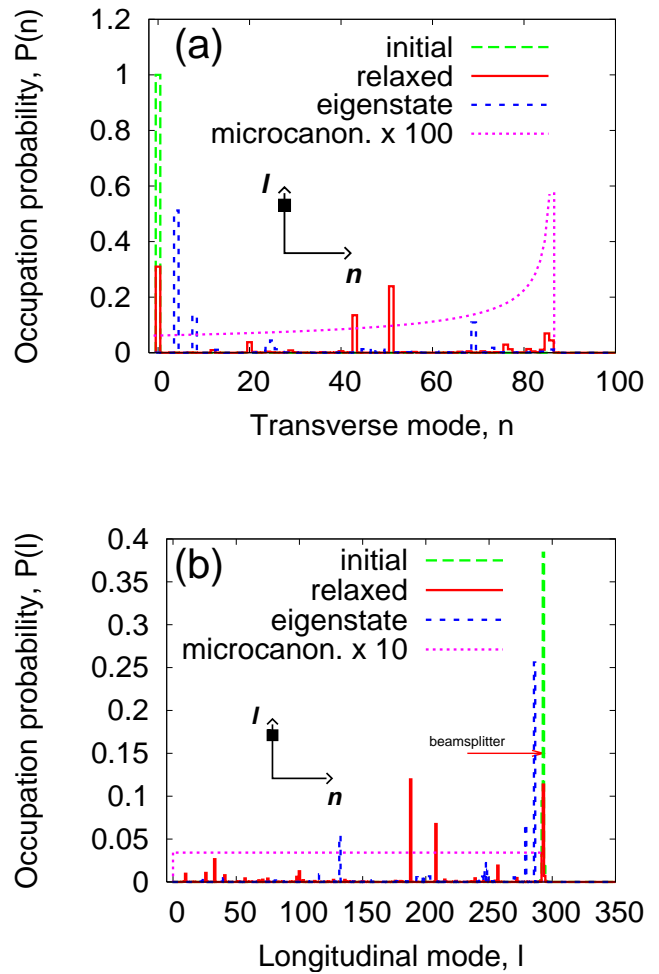}
\caption
{Distribution of the mean occupation numbers for the transverse (a) and longitudinal (b)
modes of the unperturbed Hamiltonian. The long-dashed (green) and solid (red) lines
show the initial and time-averaged distributions for an initial 
state (\protect\ref{pdn}) with $n_{0}=0$, $\delta=0.99$, and $l_{0} = 292.8$.  
Energy of the initial state is $E \equiv (2/\lambda) \hbar \omega_{\perp}\epsilon_{\alpha} = 173.2 \hbar \omega_{\perp}$.
%The state is shown as an asterix (*) on the long-dashed (green) line at the Fig.\ \ref{Fig_csie}.
The short-dashed (blue) line corresponds to the $17096$'s eigenstate 
(closest in energy to $E$ from the above, with a relative discrepancy of $10^{-5}$) of the interacting Hamiltonian.
The dotted (purple) line shows the microcanonical prediction. The rest of the parameters 
is the same as at the Fig.\ \ref{Fig_csie}.
}
\label{Fig_pdnm}
\end{figure}

For a given eigenstate, quantum expectation values of the occupation probabilities for the 
transverse (Fig.\ \ref{Fig_pdnm}a) and longitudinal 
(Fig.\ \ref{Fig_pdnm}b) modes show no convergence to their microcanonical expectation values, 
$P(n) \propto \int \! dl \, \delta(E_{nl}-E)$ and $P(l) \propto \int \! dn \, \delta(E_{nl}-E)$
respectively. 
Both distributions are heavily dominated by only a few peaks
(cf. scars in the momentum space for \v{S}eba billiards \cite{berkolaiko2003}).
%

% =======================================================================

% Relaxation

% =======================================================================

{\it Relaxation from an initial state}.--
Now we are going to address directly the ability of our system to thermalize 
from an initial state $|\psi(t\!=\!0)\rangle$. 
In this case,
the {\it infinite time average} of the quantum expectation value of an observable 
$\hat{A}$ will be given by 
$
A_{\mbox{\scriptsize relax.}} \equiv \lim_{t_{max}\to\infty} (1/t_{max}) 
\int_{0}^{t_{max}}\! \langle \psi(t) | \hat{A} | \psi(t) \rangle =
\sum_{\alpha} |\langle \alpha | \psi(t\!=\!0)|^2 \langle \alpha | \hat{A} | \alpha \rangle
$, 
where $|\alpha\rangle$ are the eigenstates of the system (see \cite{srednicki1994}).
For a quantum-chaotic system, it is expected that the above infinite time average 
coincides with the thermal prediction.  

Consider the following initial state:
\begin{equation}
%
%\langle z | \psi_{n_{0},\,\delta,\,l_{0}} \rangle
\langle z |\psi(t\!=\!0)\rangle 
\propto
\cos{\pi \zeta \over \delta}\theta \left(
 {\delta\over 2}-|\zeta |\right) \cos\left(2\pi l_{0}\zeta \right)
|n_{0}\rangle  
\,.
\label{pdn}
\end{equation}
Longitudinally, the state is represented by the ground state 
of a length 
$L\delta$ 
hard-wall box split initially by an ideal 
beamsplitter with momenta $\pm 2\pi l_{0}/L$; the box is
centered at the maximal interatomic distance.
The state is distributed among approximately  $\pi/\delta$ axial modes localized 
about $l=\pm l_{0}$.
The initial 
transverse state is limited to a single mode $n_{0}$. 
Note that for $n_{0} = 0$ and $\delta \approx 1$ the state 
is conceptually similar to the initial state used in the equilibration experiments \cite{kinoshita2006}.  

After relaxation, both transverse (Fig.\ \ref{Fig_pdnm}a) and longitudinal  (Fig.\ \ref{Fig_pdnm}b)
distributions remain very far from the equilibrium predictions. 
Fig.\ \ref{Fig_csie} shows the equilibrium value of the transverse trapping energy $U$
(see (\ref{H_2b}))
for three sequences of the initial states of the type (\ref{pdn}). 
The first two, with low transverse energy,
show an approximately -40\% deviation of the relaxed value of $U$ from 
the thermal prediction, thus being seemingly correlated with its initial, 
negative value. Note that the correlation between the final and initial values of 
observables in quantum systems is addressed in Ref.\ \cite{olshanii2009}.
Also, note that the initial states of the second sequence
have a much greater spread over the longitudinal modes than the first one; consistently, the 
energy-to-energy variation  
of the relaxed values of $U$ is less than for the first sequence. 
The third sequence, with low longitudial energy,
shows deviations from the thermal prediction that range between +50\% 
and +5\%  but never reach zero. 

Another type of the initial state
\begin{eqnarray}
&&
\langle \rho,\,z | \psi(t\!=\!0) \rangle
\propto
\cos{\pi \zeta \over
 \delta}\theta \left( {\delta\over 2}-|\zeta |\right) \cos\left(2\pi l_{0}\zeta
 \right)  
\nonumber
\\
&&
\quad
\times \left(
 e^{-\kappa _1\xi }-e^{-\kappa _2\xi }\right)  \label{pdkk}
\end{eqnarray}
allows for an additional spread over the transverse modes. 
The transverse wavefunction vanishes at the
waveguide axis. The mode occupation has a minimum at $n=0$, and the
occupation of the ground transverse mode tends to zero for $\kappa_{1} < \kappa _{2}\ll  1$.
The results for a sequence similar to the third sequence described above are 
shown at Fig.\ \ref{Fig_csie} as well. In spite of the significant initial spread 
over both transverse and longitudinal modes, the deviations from the equilibrium 
are close to the ones for the third sequence, while the energy-to-energy variations 
for the former are indeed less than for the latter.

Note that we have also successfully tested the  
convergence of the {\it instantaneous} quantum expectation values of the observables
to their infinite-time averages. 

% =======================================================================

% Summary and outlook

% =======================================================================

{\it Summary and interpretation of results}.--
In this article, we solve analytically the problem of two short-range-interacting 
atoms in a circular, transversely harmonic multimode waveguide. We assess the ability of the 
system to thermalize from an initially excited state; a broad class of initial 
states has been analyzed. We find a substantial suppression of thermalization,
even for the infinitely strong interactions. We associate this effect with 
the previously demonstrated unattainability of complete quantum chaos in
the \v{S}eba-type billiards \cite{seba1990,seba1991,albeverio1991,bogomolny2001,berkolaiko2003}.  

We conjecture that the effect of suppression of thermalization is generic for the integrable systems 
perturbed, no matter how strong, by a well localized perturbation. 
%the systems where the non-integrable perturbation 
%is localized in a single point is the full coordinate space of the system. 
The following reasoning 
applies. In a quantum chaotic system, a given eigenstate $|\alpha\rangle$ consists of a large superposition of the 
eigenstates $|\vec{n}\rangle$ of the underlying integrable system, which are drawn indiscriminately 
from the microcanonical shell. Number of principal components in such a superposition 
is a sensitive measure of the approach to a complete chaos and the subsequent thermalizability
(see, for example \cite{georgeot1997}).
The zero-point-localized perturbation,
being a particular case of a separable perturbation,
generates the eigenstates of a form 
$\langle \vec{n}|\alpha\rangle\propto 1/(E_{\alpha}-E_{\vec{n}})$ (see \cite{supplement}). 
There is always one and only one perturbed energy $E_{\alpha}$ in between any two unperturbed energies $E_{\vec{n}}$ 
(see \cite{albeverio1991}).
In addition, for the case of the infinitely strong perturbation (where the approach to 
chaos is expected to be the closest) $E_{\alpha}$ tends to the middle position between the two $E_{\vec{n}}$'s.
In this case, 
the only energy scale that remains is the distance between the 
unperturbed levels, and, as a result, the number of principal components a 
given eigenstate $|\alpha\rangle$ consists of is always of the order of unity.
This property extends to the case of finite-range interactions, whenever
the interaction range is much less then the de Broglie
 wavelength of the colliding atoms.

Our study is the first attempt to address thermalizability of a quantum system with separable interactions.

% =======================================================================

% Acknowledgments

% =======================================================================
%
\begin{acknowledgments}
We are grateful to Felix Werner for the enlightening discussions on the subject. 
This work was supported by a grant from the Office of Naval Research ({\it N00014-09-1-0502}).
\end{acknowledgments}
%
% =======================================================================
%Bibliography
% =======================================================================
%

%...............................................................
%%%%%%%%%%%%%%%%%%%%%%%%%%%%%%%%%%%%%%%%%%%%%%%%%%%%%%%%%%%%%%%%%%%%%%%%%%%%%%%%%%%%%%%%%%%%%%%%%%%%%%%%%%%%%%%%%%%%%%%%%%
%%%%%%%%%%%%%%%%%%%%%%%%%%%%%%%%%%%%%%%%%%%%%%%%%%%%%%%%%%%%%%%%%%%%%%%%%%%%%%%%%%%%%%%%%%%%%%%%%%%%%%%%%%%%%%%%%%%%%%%%%%
\end{document}

\title{Supplementary material to: Restricted Thermalization for Two Interacting Atoms in a Multimode Harmonic Waveguide}
\author{V. A. Yurovsky}
\affiliation{School of Chemistry, Tel Aviv University, 69978 Tel Aviv, Israel}
\author{M. Olshanii}
\affiliation{Department of Physics, University of Massachusetts Boston, Boston MA 02125, USA}
\date{\today}

\begin{abstract}
This supplementary material contains a detailed derivation of the eigenenergies and eigenstates for a
system consisting of two atoms in a circular,
transversely harmonic waveguide in the multimode regime.
\end{abstract}

% 05.70.Ln 	Nonequilibrium and irreversible thermodynamics 
% (see also 82.40.Bj Oscillations, chaos, and bifurcations in physical chemistry and chemical physics)
% 05.45.Mt 	Quantum chaos; semiclassical methods 
% 37.10.Gh 	Atom traps and guides
% 67.85.-d 	Ultracold gases, trapped gases (see also 03.75.-b Matter waves in quantum mechanics)
\maketitle

The waveguide system \cite{letter}, as well as the \v{S}eba billiard \cite{seba1990}, 
belongs to the class of problems where
an integrable system of Hamiltonian $\hat{H}_0$ is perturbed by a separable rank I interaction 
$V|{\cal L}\rangle\langle {\cal R}|$ (see {\it e.g.} \cite{albeverio2000}).
Eigenstates $|\alpha\rangle$ of the interacting system are solutions 
of the Schr\"odinger equation
\begin{equation}
[\hat{H}_0+V|{\cal L}\rangle\langle {\cal R}|]|\alpha\rangle=E_{\alpha}|\alpha\rangle \,,
\label{SchrSep}
\end{equation}
where $E_{\alpha}$ are the corresponding eigenenergies.
Let us expand the eigenstate $|\alpha\rangle$ over the  
eigenstates $|\vec{n}\rangle$ (of energy $E_{\vec{n}}$) 
of the non-interacting hamiltonian $\hat{H}_0$. 
The Schr\"odinger equation leads to the following equations for the expansion coefficients
\begin{equation}
(E_{\alpha}-E_{\vec{n}})\langle \vec{n}|\alpha\rangle =
\langle \vec{n} |{\cal L} \rangle\langle {\cal R}|\alpha\rangle
\propto \langle \vec{n} |{\cal L} \rangle ,
\end{equation}
where the omitted factor in the second equality is independent of $\vec{n}$.

As a result, any eigenfunction  $|\alpha\rangle$
functionally coincides with the Green function of the 
non-interacting Hamiltonian taken at the energy of the eigenstate $E_{\alpha}$,
\begin{equation}
|\alpha\rangle \propto \sum\limits_{\vec{n}}
{|\vec{n}\rangle \langle \vec{n} |{\cal L} \rangle
 \over E_{\alpha}-E_{\vec{n}}}.
\label{psiGreen}
\end{equation}
The omitted factor is determined by the normalization conditions.
 
Substitution of this expression to the Schr\"odinger equation (\ref{SchrSep}) leads to
 the following eigenenergy equation: 
\begin{equation}
\sum\limits_{\vec{n}}
{\langle {\cal R}|\vec{n}\rangle \langle \vec{n} |{\cal L} \rangle
 \over E_{\alpha}-E_{\vec{n}}} =\frac{1}{V}.
\label{EnerSep}
\end{equation}
Similar
expressions 
were obtained in Refs.\
\cite{seba1990,albeverio1991} for the case of the \v{S}eba billiard and its generalizations.

In the case of the relative motion of two short-range-interacting 
atoms in a circular, transversely harmonic waveguide, the derivation
uses ideas of the analogous model involving an infinite waveguide 
\cite{olshanii1998,moore2004,yurovsky2008b}. The unperturbed Hamiltonian here is given by Eq.\ (1) in \cite{letter} 
and the Fermi-Huang interaction (see Eq.\ (2) in \cite{letter}) can be written in the separable form \cite{albeverio2000}  
with the interaction strength and formfactors given by
\begin{equation}
V={2\pi\hbar^2 a_{s}\over \mu},
\quad
|{\cal L}\rangle=\delta_{3}({\bm r}),
\quad
\langle {\cal R}|=\delta_{3}({\bm r}) {\partial\over\partial r}(r\,\cdot ).
\end{equation}
In what follows, we will restrict the Hilbert space to 
the states of zero $z$-component of the angular momentum and even under the $z \leftrightarrow -z$ reflection; 
the interaction has no effect on the rest of the Hilbert space. 
The non-interacting eigenstates are products of the transverse and longitudial wavefunctions.
The transverse two-dimensional zero-angular-momentum 
harmonic wavefunctions, 
\begin{equation}
\langle \rho |n\rangle ={1\over \sqrt{\pi }a{ } _{\perp }}L^{\left(
 0\right) }_{n}\left( \xi \right) \exp\left( -\xi /2\right)  ,
 \label{wf_trans}
\end{equation}
labeled by the quantum number $n \geq 0$,
are expressed in terms of the Legendre polynomials, $ L^{\left(
 0\right) }_{n}\left( \xi \right)$, where  $\xi =\left( \rho
/a_{\perp }\right) ^{2}$ and $a_{\perp }=\left( \hbar/\mu \omega _{\perp }\right) ^{1/2}$
is the transverse oscillator range.
The longitudial wavefunctions, labeled by  $l\geq 0$, are the symmetric plane waves
 satisfying periodic conditions with the period $L$: 
\begin{equation}
\langle z|k\rangle =\left( 2/L\right) ^{1/2}\cos 2\pi k\zeta  ,\qquad
 \langle z|0\rangle =L^{-1/2} ,
\label{wf_axial}
\end{equation} 
where $\zeta \equiv  z/L-1/2$.
 The unperturbed spectrum is therefore given by
\begin{equation}
E_{nl} = 2 \hbar \omega_{\perp} n + \hbar^2 (2\pi l/L)^2/(2\mu)
\label{E0_2b}
\end{equation}

Substituting the above non-interacting eigenstates and eigenenergies to Eq.\ (\ref{psiGreen}) and using
Eq.\ (1.445.8) in \cite{gradshteyn} for summation over $l$, 
one obtains Eq.\ (3) in \cite{letter} for the interacting eigenstates.
The normalization factor $C_{\alpha} $ is determined by the condition
\begin{equation}
2\pi\int\limits_0^L dz\int\limits_0^{\infty} \rho^2 d\rho \langle \alpha'|\rho,z\rangle
\langle \rho,z|\alpha \rangle =\delta_{\alpha\alpha'}.
\end{equation}
Using orthogonality of the Legendre polynomials it can be expressed as
\begin{eqnarray}
C_{\alpha} ={2\over a_{\perp }\sqrt{\pi L}}
\biggl\lbrack \sum\limits^{\infty }_{n=0}\biggl
({\cot\sqrt{\epsilon_{\alpha} -\lambda n}\over \left( \epsilon_{\alpha} -\lambda
 n\right) { } ^{3/2}} \nonumber
 \\
+{1\over \left( \epsilon_{\alpha} -\lambda n\right) \sin^{2}\sqrt{\epsilon_{\alpha}
 -\lambda n}}\biggr)\biggr\rbrack ^{-1/2} , \label{Ceps}
\end{eqnarray}
where the rescaled energy $\epsilon_{\alpha}$ is given by $\epsilon_{\alpha}\equiv\lambda E_{\alpha}/\left(
 2\hbar\omega_{\perp }\right)$ and $\lambda \equiv\left( L
/a_{\perp }\right) ^{2}$  is the aspect ratio.

For the Fermi-Huang interaction the eigenenergy equation (\ref{EnerSep})
attains the form
\begin{equation}
{\partial \over \partial r}\left [ r\sum\limits^{}_{n,l}{\langle \rho
|n\rangle \langle z|l\rangle \langle n|0\rangle \langle 0|0\rangle
 \over E-E_{nl}}\right ]_{r=0}={1\over V}
 \end{equation}
In the limit $r\to 0$, the eigenstate is spherically symmetric. This allows us 
to deal with the $z$-axis only (see \cite{moore2004}).
Substitution of the noninteracting wavefunctions (\ref{wf_trans}) and (\ref{wf_axial})
and energies (\ref{E0_2b}) with the subsequent summation over $l$ leads to
\begin{equation}
\sqrt{\lambda}{\partial \over \partial z}\left [ z \sum\limits^{\infty
 }_{n=0}{\cos\left( 2\sqrt{\epsilon _{\alpha}-\lambda n}(z/L-1/2) \right)
 \over \sqrt{\epsilon _{\alpha}-\lambda n}\sin\sqrt{\epsilon _{\alpha}-\lambda
 n}}\right ]_{z=0}={a_s \over a_{\perp}}.
\end{equation}
This sum contains both the regular part and the irregular one, the latter being proportional
to $z^{-1}$. The regular part can be extracted using the identity
\begin{eqnarray}
\lim_{z\rightarrow 0}\left\lbrack \sum\limits^{\infty }_{n=0}
(\lambda n-\epsilon)^{-1/2}\exp(-2\sqrt{\lambda n-\epsilon}z/L)-
{L\over \lambda z}\right \rbrack \nonumber
\\
=-{1\over \sqrt{\lambda }}\zeta \left( {1\over 2},-{\epsilon\over \lambda}\right)
\end{eqnarray} 
(see \cite{moore2004})
where $\zeta \left( \nu ,\alpha \right) $ is the Hurwitz zeta
 function (see, e. g., \cite{gradshteyn,yurovsky2008b}.
Finally we arrive at the
transcendental equation for the eigenenergies (4) in \cite{letter}.
%V
The summands in the sums Eqs.\ (3) and (4) in \cite{letter} and
in Eq.\ (\ref{Ceps}) 
decay exponentially with $n$, leading to the fast converging series.
Note that the imaginary parts of the two terms in the left hand side of Eq.\ (4)
in \cite{letter} cancel each other automatically.

%...............................................................
%%%%%%%%%%%%%%%%%%%%%%%%%%%%%%%%%%%%%%%%%%%%%%%%%%%%%%%%%%%%%%%%%%%%%%%%%%%%%%%%%%%%%%%%%%%%%%%%%%%%%%%%%%%%%%%%%%%%%%%%%
%%%%%%%%%%%%%%%%%%%%%%%%%%%%%%%%%%%%%%%%%%%%%%%%%%%%%%%%%%%%%%%%%%%%%%%%%%%%%%%%%%%%%%%%%%%%%%%%%%%%%%%%%%%%%%%%%%%%%%%%%%